\documentclass[amsmath,amssymb,preprint,showpacs,showkeys,byrevtex,tightenlines]{revtex4-1}
\usepackage{graphicx}
\usepackage{bm}
\newcommand*{\fh}[1]{#1\nobreak\discretionary{}%
{\hbox{$\mathsurround=0pt #1$}}{}}

\begin{document}

\title{Spatial distribution of atoms in the field of the criss-cross
  standing bichromatic light waves}



\begin{abstract}
  We show that properly detuning the carrier frequency of each of the
  criss-cross bichromatic waves from the transition frequency in the
  atom, it is possible to form a two-dimensional trap for atoms if the
  intensity of the waves is sufficiently large. For zero and near zero
  initial phases of waves, and also for $\pi$ and near $\pi$ phase
  shift between criss-cross waves a dynamic spatial structure of
  square cells with the side $\lambda /\sqrt{2}$ is formed. Numerical
  simulations are carried out for sodium atoms.
\end{abstract}

\author{V. I. Romanenko} \affiliation{Institute of Physics,
  Nat. Acad. of Sci. of Ukraine (46, Nauky Ave., Kyiv 03680, Ukraine)}
\author{N. V. Kornilovska} \affiliation{Kherson National Technical
  University (24, Berislavske shosse,
  Kherson  73008, Ukraine)}
\author{O. G. Udovytska} \affiliation{Institute
  of Physics, Nat. Acad. of Sci. of Ukraine (46, Nauky Ave., Kyiv
  03680,\,\! Ukraine)}
\author{L. P. Yatsenko} \affiliation{Institute of
  Physics, Nat. Acad. of Sci. of Ukraine (46, Nauky Ave., Kyiv 03680,\,\!
  Ukraine)}

\maketitle{}

\tableofcontents{}

\section{Introduction}
\label{sec:introduction}
Various aspects of the interaction of atoms with bichromatic standing
waves (a pair of monochromatic standing waves with different
frequencies), which can also be considered as counter-propagating
bichromatic waves or counter-propagating amplitude-modulated waves,
are have been studied already for three decades.  The physical basis
of the interaction of atoms with bichromatic waves and numerous
experimental work are analyzed, in particular, in the
books~\cite{Metcalf,Negriyko}, and in the recent
review~\cite{Metcalf-2017}.  The fact that the light pressure force on
the atom in the field of a standing bichromatic wave can significantly
exceed the force of light pressure in the field of a travelling
monochromatic
wave~\cite{Voitsekhovich88,Kazantsev-Krasnov,Voitsekhovich89,Grimm,Ovchinnikov,Soding}
is essential for the efficient manipulation of atomic
beams~\cite{Soding,Williams,Cashen-Metcalf}.  The field of bichromatic
waves, as it turned out, can be used to cool atoms and molecules
without the participation of spontaneous emission
\cite{Corder,Corder-JOSA}. This new phenomenon, predicted
in~\cite{Metcalf-2008}, will probably enlarge our potential of
controlling the motion of molecules by the laser radiation.  Another
direction of research concerns the formation of traps for atoms only
by laser radiation, without additional fields (for example, magnetic
field in magneto-trap~\cite{Metcalf}).  The first proposition of such
traps was based on the interaction of atoms with the sequences of the
counter-propagating light
pulses~\cite{Freegarde,Goepfert,Romanenko-JOPB,Romanenko-PRA}, then it
turned out that the traps can be formed counter-propagating
bichromatic~\cite{Romanenko2016}, stochastic~\cite {Romanenko-OC} and
frequency-modulated~\cite{Romanenko-EPJD} waves.  In addition, the laser 
field that confines atoms in the trap can also cools them.

We have considered the interaction of two-level atoms with two
orthogonal linearly polarized standing bichromatic waves and found the
conditions, for which these waves form a two-dimensional trap for
atoms (one-dimensional trap was analyzed in~\cite{Romanenko2016}). We
have also found the shape and parameters of a grating of atoms, which
is formed as a result of their interaction with the field. To describe
the motion of atoms in the field, we use the Newton's laws, and the
evolution of atomic states we describe by the Monte Carlo
wave-function method~\cite{Mol93}. Numerical simulation is carried out
for sodium atoms.

The paper is oraganized as follows. In the next section we describe
the interaction of the atom with the field, the third section provides
the basic equations, the fourth section briefly describes the Monte
Carlo wave function method, in the fifth section we ex[lain] the
procedure of numerical calculation, the obtained results and their
discussion are given in the sixth section, short conclusions are
formulated at the end of the paper.

\section{Interaction of the atom with the field}
\label{sec:interaction}
\begin{figure}[h]
\centerline{\includegraphics[]{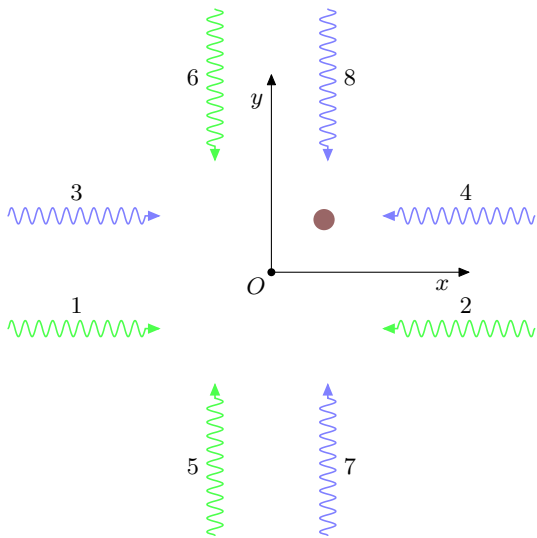}}
\caption{Schematic interaction of the atom with the field. At the
  point $O$ the phase difference between the opposite waves is
  zero. Atom is marked with a circle.}
\label{fig-1}
\end{figure}

We consider the atom moving in the field of two criss-cross
bichromatic standing waves, each of which is formed by pairs of
collinear standing waves. In turn, each of these standing waves can be
considered as two counter-propagating monochromatic waves, as shown in
Fig.~\ref{fig-1}.

We consider the interaction of atoms with the bichromatic field of
standing waves near the point $O$, where the wave antinodes coincide,
that is, the difference between the phases of counter-propagating
waves here is zero. In the field of one bichromatic wave, a slight
deviation from this point leads to the light pressure force acting on
the atom. This force is proportional to to the sinus the phase
difference between the standing monochromatic waves that form the
standing bichromatic wave~\cite{Voitsekhovich88, Yatsenko}. Since this
phase difference linearly depends on the coordinates of the atom, a
bichromatic standing light wave can form a one-dimensional trap for
atoms~\cite{Voitsekhovich89, Romanenko2016}. Obviously, one can expect
that the field of two criss-cross bichromatic waves can form a
two-dimensional trap for atoms.

We should note that the interaction of an atom with two criss-cross
standing bichromatic waves qualitatively differs from the interaction
with one wave.  In addition to the absorption of a photon from one
travelling bichromatic wave with the following emission of a photon
into the oncoming travelling bichromatic wave, there may be the
absorption of a photon from one travelling bichromatic wave with
subsequent emission of a photon into the orthogonal travelling
bichromatic wave. Thus, the result of the atom's interaction with the
field of two standing bichromatic waves can not be considered as the
sum of the atomic interactions with each of the standing bichromatic
waves.

\section{The main equations}
The interaction of an atom with a field of cris-scross bichromatic
waves can be concidered as its interaction with the field of eight
travelling monochromatic waves, as shown in Fig.~\ref{fig-1}. Spatial
and time dependence of the field strength of each of them can be
described by the equations
\begin{eqnarray}
  \mathbf{E}_{n=1,2,3,4}&=&\mathbf{e}E_{0n}
  \cos\left(i\omega_{n} t\mp{}ik_{n}x+i\varphi_{n}\right),
\label{eq:E-monox}\\
  \mathbf{E}_{n=5,6,7,8}&=&\mathbf{e}E_{0n}
  \cos\left(i\omega_{n} t\mp{}ik_{n}y+i\varphi_{n}\right),
\label{eq:E-monoy}
\end{eqnarray}
where the sign `` $-$ '' corresponds to an odd $n$, ``$+$'' corresponds to an
even $n$, $\mathbf{e}$ is the unit polarization vector, $\omega_{n}$
is the frequency of $n$-th monochromatic wave.  Note that we chose the
same direction of the polarization vectors of all the waves.
The condition of forming four standing waves by these eight travelling waves is 
\begin{equation}
  \label{eq:frequencies}
  \omega_{2n}=\omega_{2n-1}, \quad n=1,2,3,4.
\end{equation}
For the sake of simplicity we consider the standing waves with equal
amplitudes,
\begin{equation}
  E_{0n}=E_{0}, \quad n=1,\ldots,8.\label{eq:standing_E}
\end{equation}
We define the detunings $\delta_n$ of the frequencies $\omega_n$ from
the atomic transition frequency $\omega_0$ as
\begin{equation}
  \label{eq:detunings}
  \delta_{n}=\omega_{0}-\omega_{n},\quad n=1,\ldots,8,
\end{equation}
as far as the frequencies of opposing monochromatic waves that form
each of the four standing waves are the same~\eqref{eq:frequencies}.
We assume that the frequency difference between the monochromatic
standing waves that form the bichromatic standing waves, are also the
same for both of them:
\begin{equation}
  \label{eq:mod}
  \omega_{1}-\omega_{3}=\omega_{5}-\omega_{7}=\Omega>0.
\end{equation}
For further simplification, we assume equal mean frequencies of the
criss-cross waves, i.e.
\begin{equation}
  \label{eq:carr}
  \frac{\omega_{3}+\omega_{1}}{2}=\frac{\omega_{7}+\omega_{5}}{2}=\omega.
\end{equation}
From these equations, it is easy to express the frequencies of all
monochromatic waves through the mean frequency $\omega$ and the
frequency difference $\Omega$ of the waves:
\begin{equation}
  \label{eq:om13}
\omega_{1}=\omega_{5}=\omega+\frac{\Omega}{2},\quad
  \omega_{3}=\omega_{7}=\omega-\frac{\Omega}{2}.
\end{equation}
Introducing the detuning
\begin{equation}
  \label{eq:delta}
\delta=\omega_{0}-\omega
\end{equation}
of the mean frequency of bichromatic waves from the transition
frequency of the atom and, taking into account~\eqref{eq:om13}
and~\eqref{eq:detunings}, we obtain:
\begin{equation}
  \label{eq:d13}
\delta_{1}=\delta_{5}=\delta-\frac{\Omega}{2},\quad
\delta_{3}=\delta_{7}=\delta+\frac{\Omega}{2}.
\end{equation}
The field of eight travelling monochromatic waves can be written as a
field of four standing waves, the field strength of which has the form
\begin{eqnarray}
  \mathbf{E}_{12}&=&\mathbf{E}_{1}+\mathbf{E}_{2}=
                   2E_{0}\mathbf{e}\cos\left[\omega_{1}t+\tfrac{1}{2}\left(\varphi_{1}+\varphi_{2}\right)\right]
\cos\left[k_{1}x+\tfrac{1}{2}\left(\varphi_{2}-\varphi_{1}\right)\right],
                   \label{eq:12}\\
  \mathbf{E}_{34}&=&\mathbf{E}_{3}+\mathbf{E}_{4}=
                   2E_{0}\mathbf{e}\cos\left[\omega_{3}t+\tfrac{1}{2}\left(\varphi_{3}+\varphi_{4}\right)\right]
\cos\left[k_{3}x+\tfrac{1}{2}\left(\varphi_{4}-\varphi_{3}\right)\right],
                   \label{eq:34}\\
  \mathbf{E}_{56}&=&\mathbf{E}_{5}+\mathbf{E}_{6}=
                   2E_{0}\mathbf{e}\cos\left[\omega_{1}t+\tfrac{1}{2}\left(\varphi_{5}+\varphi_{6}\right)\right]
\cos\left[k_{1}y+\tfrac{1}{2}\left(\varphi_{6}-\varphi_{5}\right)\right],
                   \label{eq:56}\\
  \mathbf{E}_{78}&=&\mathbf{E}_{3}+\mathbf{E}_{4}=
                   2E_{0}\mathbf{e}\cos\left[\omega_{3}t+\tfrac{1}{2}\left(\varphi_{7}+\varphi_{8}\right)\right]
\cos\left[k_{3}y+\tfrac{1}{2}\left(\varphi_{8}-\varphi_{7}\right)\right],
                   \label{eq:78}
\end{eqnarray}
and the total field acting on the atom can be written as
\begin{equation}
  \mathbf{E}=\mathbf{E}_{12} +\mathbf{E}_{34}+\mathbf{E}_{56} +\mathbf{E}_{78}.
\label{eq:sumst}
\end{equation}

Since the first studies of the mechanical action of the bichromatic
waves on atoms~\cite{Voitsekhovich88,Voitsekhovich89}, to calculate
the light pressure on atoms the representation of the field in the
form of a superposition of the counter-propagating amplitude-modulated
waves is also used. This makes it possible to draw an analogy between
the field of the bichromatic waves and the field of the sequence of
counter-propagating light pulses. This analogy helps to understand why
the light pressure force in the field of the bichromatic waves
significantly exceeds the light pressure force on an atom in a single
travelling wave~\cite{Kazantsev}. In the case of two criss-cross
standing waves these counter-propagating waves read
\begin{eqnarray}
  \mathbf{E}_{13}&=&
                   \mathbf{E}_{1}+\mathbf{E}_{3}=
                   2E_{0}\mathbf{e}\cos\left[\omega t-kx+
                   \tfrac{1}{2}(\varphi_{1}+\varphi_{3})\right]
\cos\left[\tfrac{1}{2}\Omega t-
                   \tfrac{1}{2}\Delta kx+\tfrac{1}{2}(\varphi_{3}-\varphi_{1})\right],
                   \label{eq:ampl_i_iii}\\
  \mathbf{E}_{24}&=&\mathbf{E}_{2}+\mathbf{E}_{4}=
                   2E_{0}\mathbf{e}\cos\left[\omega t+kx+
                   \tfrac{1}{2}(\varphi_{2}+\varphi_{4})\right]
\cos\left[\tfrac{1}{2}\Omega t+
                   \tfrac{1}{2}\Delta kx+\tfrac{1}{2}(\varphi_{4}-\varphi_{2})\right],
                   \label{eq:ampl_ii_iv}\\
  \mathbf{E}_{57}&=&
                   \mathbf{E}_{1}+\mathbf{E}_{3}=
                   2E_{0}\mathbf{e}\cos\left[\omega t-ky+
                   \tfrac{1}{2}(\varphi_{5}+\varphi_{7})\right]
\cos\left[\tfrac{1}{2}\Omega t-
                   \tfrac{1}{2}\Delta ky+\tfrac{1}{2}(\varphi_{7}-\varphi_{5})\right],
                   \label{eq:ampl_v_vii}\\
  \mathbf{E}_{68}&=&\mathbf{E}_{2}+\mathbf{E}_{4}=
                   2E_{0}\mathbf{e}\cos\left[\omega t+ky+
                   \tfrac{1}{2}(\varphi_{6}+\varphi_{8})\right]
\cos\left[\tfrac{1}{2}\Omega t+
                   \tfrac{1}{2}\Delta ky+\tfrac{1}{2}(\varphi_{8}-\varphi_{6})\right],
                   \label{eq:ampl_vii_viii}
\end{eqnarray}
where
\begin{eqnarray}
  k&=&\frac{1}{2}\left(k_{1}+k_{3}\right)=\frac{1}{2}\left(k_{2}+k_{4}\right),
     \label{eq:not-i} \\
  \Delta k&=&k_{3}-k_{1}=k_{4}-k_{2}=\Omega/c.
            \label{eq:not-ii} 
\end{eqnarray}
The change of the initial time or the reference frame ($x$ and $y$) is
obviously equivalent to the change of initial phases $\varphi_{n}$
($n=1,\ldots,8$).  Since $\Omega\ll\omega$ and $\Delta k\ll k $
changing the initial time, one can make two initial phases equal to
zero, and changing the origin of the reference frame one can make four
initial phases equal to zero (or their linear combinations).

Projections of the force $\mathbf{F}$ acting on the atom on $Ox$ and
$Oy$ axes are~\cite{Minogin, Metcalf}
\begin{eqnarray}
  F_{x}=(\varrho_{12}\mathbf{d}_{21}+\varrho_{21}\mathbf{d}_{12})\frac{\partial \bm{\mathrm{E}}}{\partial x},
  \label{eq:Fx}\\
 F_{y}=(\varrho_{12}\mathbf{d}_{21}+\varrho_{21}\mathbf{d}_{12})\frac{\partial \bm{\mathrm{E}}}{\partial y},
  \label{eq:Fy}
\end{eqnarray}
where $\mathbf{d}_{ij}$ ($i, j = 1,2$) are the matrix elements of the
dipole moment, $\varrho_{ij}$ are the elements of the density matrix
$\varrho$. The motion of an atom under the action of this force is
described by the second Newton's law
\begin{equation}
  \ddot{\mathbf{r}}=\mathbf{F}/m,\label{eq:ma}
\end{equation}
where $ m $ is the mass of the atom, $\mathbf{r}$ is the radius vector
of the atom with coordinates $ x $ and~$ y $.

We find the density matrix using the probability amplitudes
$ c_ {1} $, $ c_ {2} $ of the occupation of the ground
$ \left | 1 \right \rangle $ and excited $ \left | 2 \right \rangle $
states by the atom:
\begin{equation}
\varrho_{12}=c_{1}c_{2}^{*} e^{i\omega_{0}t},\quad\varrho_{21} =c_{2}c_{1}^{*} e^{-i\omega_{0}t}.
\label{eq:cc}
\end{equation} 
The state vector
\begin{equation}
\left| \psi\right\rangle=c_{1}\left| 1\right\rangle+c_{2}e^{-i\omega_{0}t}\left| 2\right\rangle,
\label{eq:psi}
\end{equation}
of the atom is calculated from the Schr\"o{}dinger equation
\begin{equation}
i\hbar\frac{d}{dt}\left| \psi\right\rangle=H\left| \psi\right\rangle.
\label{eq:Schrod}
\end{equation}
by modeling the state vector using the Monte Carlo wave-function
method~\cite {Mol93}, which takes into account the possibility of
spontaneous emission of light by the atom. Contrary to computations
based on the density matrix, this method of calculation allows us to
simulate the trajectory of the motion of a single atom.

\section[Modelling of the state vector by Monte Carlo wave function
method]{Modelling of the state vector by Monte Carlo wave function
  method}
The Monte Carlo wave-function method for modeling of the state
vector~\cite{Mol93} peivides the numerical solution of the
Schr\"o{}dinger equation~\eqref{eq:Schrod} with the possibility of
spontaneous emission of light by an atom. The spontaneous emission is
taken into account by the relaxation term
\begin{equation}
  H_{rel}= -\frac{i\hbar}{2}\gamma|2\rangle\langle{}2|,
  \label{eq:Hrel}
\end{equation}
where $\gamma$ is the spontaneous emission rate, in Hamiltonian
\begin{equation}
  {{H}}=H_{0}+H_{int}+H_{rel}.
  \label{eq:Ham}
\end{equation}
Here
\begin{equation}
  H_{0}= \hbar\omega_{0}|2\rangle\langle2|
  \label{eq:H0}
\end{equation}
describes an atom in the absence of a field and relaxation, and the term
\begin{equation}
    H_{int}= -\bm{\mathrm{d}}_{12}|1\rangle\langle2|\bm{\mathrm{E}}(t)-\bm{\mathrm{d}}_{21}|2\rangle\langle1|\bm{\mathrm{E}}(t)
\label{eq:Hint}
\end{equation}
is responsible for the interaction of the atom with the field.

Since the Hamiltonian (\ref{eq:Ham}) is not Hermitian, we normalize
the state vector \eqref{eq:psi} after each small step in time. The
value of the deviation of the vector norm from the unit is used in
this case to simulate the process of spontaneous photon emission.
With the increase of the deviation, the probability of spontaneous
emission increases. The spontaneous emission leads to a quantum jump
from an excited state $ \left | 2 \right \rangle $ into the ground
state $ \left | 1 \right \rangle$.  Here we use the Monte Carlo
wave-function method of the first order of accuracy, described in
\cite{Mol93}. Methods of the second and fourth order accuracy are
considered in \cite{Steinbach}.

Let the atom at the time $ t $ be described by the state vector
$ | \psi (t) \rangle $. We find the state vector
$ | \psi (t + \Delta  t) \rangle $ at the time $ t + \Delta {} t $
in two steps.

\textbf{1}. Integrating Schr\"odinger equation (\ref{eq:Schrod}) we
find the state vector after a sufficiently small~$ \Delta {} t $:
\begin{equation}
  |\psi^{(1)}(t+\Delta{}t)\rangle=\left(1-\frac{i\Delta{}t}{\hbar}{{H}}\right)|\psi(t)\rangle.
  \label{eq:phiI}
\end{equation}
After this step in time, the norm of the state vector equals
\begin{equation}
  \langle\psi^{(1)}(t+\Delta{}t)|\psi^{(1)}(t+\Delta{}t)\rangle=1-\Delta{}P,
  \label{eq:phiIN}
\end{equation} 
where
\begin{equation}
  \Delta{}P=\frac{i\Delta{}t}{\hbar}\langle\psi(t)|H-H^{+}|\psi(t)\rangle=\gamma|\Delta{}t|c_{2}|^{2}.
\label{eq:dPpsi}
\end{equation}

\textbf{2}.  At the second step, we check whether there was a quantum
jump (which accompanies spontaneous emission) during the integration
time $ \Delta t $. For this purpose we compare the value of the random
variable $ \epsilon $, which is uniformly distributed between zero and
one, with $ \Delta {} P $. If $ \epsilon <\Delta {} P $, the quantum
jump occurs, the atom passes into the state $ | 1 \rangle $ and the
state vector becomes
\begin{equation}
  |\psi(t+\Delta{}t)\rangle=|1\rangle, \quad \epsilon<\Delta{}P.
\end{equation}
If $ \epsilon> \Delta {} P $ (in most cases, since
$ \Delta {} P \ll1 $), the jump does not occur and state vector
\eqref{eq:phiI} obtained at the first stage is normalized:
\begin{equation}
  |\psi(t+\Delta{}t)\rangle=\frac{|\psi^{(1)}(t+\Delta{}t)\rangle}{\sqrt{1-\Delta{}P}}, \quad \Delta{}P<\epsilon.
\end{equation}

The direction of propagation of a photon after spontaneous emission we
also simulate by the Monte Carlo method, assuming that the propagation
directions along the axes $ Ox $ and $ Oy $ are equally probable
(details are given in the fifth section).

Although equation \eqref{eq:phiI} gives a formal solution to the
Schr\"odinger equation, it is more convenient to use the equations for
the probability amplitudes $ c_ {1} $, $ c_ {2} $ of populations of
states $ \left | 1 \right \rangle $ and $ \left | 2 \right \rangle $,
and, having found them, get the state vector \eqref{eq:psi}. The
equations for the amplitudes follow from the Schr\"odinger equation
\eqref{eq:Schrod} and are equivalent to it.

Substitution of~(\ref{eq:psi}), (\ref{eq:Ham}) in (\ref{eq:Schrod})
gives
\begin{eqnarray}
\label{eq:ci}
  i\hbar\frac{d}{dt}c_{1}&=&-\bm{\mathrm{d}}_{12}
  \bm{\mathrm{E}}c_{2}e^{-i\omega_{0}t},\\
  i\hbar\frac{d}{dt}c_{2}&=&-\bm{\mathrm{d}}_{21}
  \bm{\mathrm{E}}c_{1}e^{i\omega_{0}t}-\frac{1}{2}\gamma c_{2}.
\label{eq:cii}
\end{eqnarray}
In the rotating wave approximation (neglecting the exponential terms
$ \sim {} e ^ {\pm2i \omega_ {0} t} $) \cite{Shore} from the
equations (\ref{eq:ci}), (\ref{eq:cii}) and taking into account
\begin{equation}
  \mathbf{E}=  \sum\limits_{n=1}^{8}\mathbf{E}_{n}
  \label{eq:E}
\end{equation}
we get
\begin{eqnarray}
    \frac{d}{dt}c_{1}&=&-\frac{i}{2}\Omega_{0}\sum
    \limits_{n=1}^{4}e^{(-1)^{n}ik_{n}x-i\delta_{n} t+i\varphi_{n}}c_{2}
-\frac{i}{2}\Omega_{0}\sum
    \limits_{n=5}^{8}e^{(-1)^{n}ik_{n}y-i\delta_{n} t+i\varphi_{n}}c_{2},
 \label{eq:dci}\\
    \frac{d}{dt}c_{2}&=&-\frac{i}{2}\Omega_{0}^{*}
    \sum\limits_{n=1}^{4}e^{(-1)^{n+1}ik_{n}x+i\delta_{n} t-i\varphi_{n}}c_{1} 
 -\frac{i}{2}\Omega_{0}^{*}
    \sum\limits_{n=5}^{8}e^{(-1)^{n+1}ik_{n}y+i\delta_{n} t-i\varphi_{n}}c_{1} 
-\frac{1}{2}\gamma c_{2},
  \label{eq:dcii}
\end{eqnarray} 
where the Rabi frequency of the monochromatic waves
\begin{equation}
  \Omega_{0}=-\bm{\mathrm{d}}_{12}\bm{\mathrm{e}}E_{0}/\hbar,
  \label{eq:Rabi}
\end{equation}
is introduced, which, without loss of the generality, we consider to be
real \cite{Shore}.

Knowing the probability amplitudes of $ c_{1}$, $ c_{2}$, one can
calculate the light pressure force acting on the atom using
(\ref{eq:Fx}), (\ref{eq:Fy}) and \eqref{eq:cc}.  After the averaging
of the expressions for the force (\ref{eq:Fx}), (\ref{eq:Fy}) over
time in an interval much greater than the time of fast oscillations
with the characteristic time $ 2 \pi / \omega_{0} $ and at the same
time, small enough that the average light pressure force practically
does not depend on the time of averaging, we find
\begin{eqnarray}
  F_{x}&= &  \hbar{}\sum\limits_{n=1}^{4}(-1)^{n+1}k_{n}
            \mathop{\mathrm{Im}}\bigl[c_{1}c_{2}^{*}\Omega_{n}^{*}e^{i\delta_{n} t-i\varphi_{n}} 
            e^{i(-1)^{n+1}k_{n}x}
            \bigr]\left({|c_{1}|^{2}+|c_{2}|^{2}} \right)^{-1}.
            \label{eq:FFx}\\
  F_{y}&= &  \hbar{}\sum\limits_{n=5}^{8}(-1)^{n+1}k_{n}
            \mathop{\mathrm{Im}}\bigl[c_{1}c_{2}^{*}\Omega_{n}^{*}e^{i\delta_{n} t-i\varphi_{n}} 
            e^{i(-1)^{n+1}k_{n}y}
            \bigr]\left({|c_{1}|^{2}+|c_{2}|^{2}} \right)^{-1}.
\label{eq:FFy}
\end{eqnarray}
Now we can describe the motion of an atom by integrating
simultaneously the equation \eqref {eq:ma}, \eqref {eq:dci},
\eqref{eq:dcii} and taking into account the expressions
\eqref{eq:FFx}, \eqref{eq:FFy} for the projections $ F_{x} $,
$ F_{y} $ of the light pressure force on the abscissa and ordinate
axes.

The normalization in \eqref{eq:FFx}, \eqref{eq:FFy} is required for
combining integration of the Schr\"odinger equation and equations of
motion by the fourth method order accuracy with the Monte Carlo wave
function method of the first order because the intermediate time
points are used.

\section{Numerical calculation routine}

To simulate the motion of an atom, we simultaneously solve the Newton
equation (\ref{eq:ma}) with the force (\ref{eq:FFx}), (\ref{eq:FFy})
and the equations (\ref{eq:dci}), (\ref{eq:dcii}) for the amplitudes
of the probability of states of the atom. In addition, we simulate the
momentum change of the atom due to the spontaneous emission of light
by the atom and due to absorption and stimulated emission
fluctuations, which also results in the to atomic momentum
fluctuations. In our calculations, for simplicity, we assume that the
spontaneous emission of a photon by an atom leads to the change
$\hbar k$ in the atom's momentum with the same probability in the
positive and negative directions of the axes $ Ox $ and $ Oy $.

The stochastic nature of the spontaneous emission of light by an atom
causes the atom's diffusion in the momentum space, the so-called
``momentum diffusion''.  In the field of low-intensity laser
radiation, when the population of the excited state is insignificant,
the light pressure force and the momentum diffusion coefficient are
equal to the sum of the corresponding quantities for each of the
counter-propagating waves \cite{Mol91}.  We used this approximation
earlier to simulate the momentum diffusion in the field of
counter-propagating light pulses of low intensity
\cite{Romanenko-PRA}, as well as for estimation of momentum diffusion
in the field of bichromatic \cite{Romanenko2016}, stochastic
\cite{Romanenko-OC,Romanenko2018} and frequency-modulated
\cite{Romanenko-EPJD} light waves.

We simulate the change in the atomic momentum due to the recoil effect
of the spontaneous emission by modeling the direction of the photon
propagation. For example, we may divide the interval from 0 to 1 into
four identical intervals, each of which corresponds to the radiation
of a photon along the positive or negative axes direction $ Ox $,
$ Oy $ and check, if the uniformely distributed between 0 and 1 random
variable gets into one of four intervals.

Calculation of the momentum fluctuation due to the absorption and
stimulated emission fluctuations requires additional attention.  Let's
consider the atomic momentum fluctuations in the field of a travelling
monochromatic waves. Let $ \theta $ be the angle between the direction
of the spontaneously emitted photon and axis $ Ox $, along which the
light wave propagates, $ \langle N_{s} \rangle $ is the average number
of spontaneously emitted photons. Then, assuming the scattering of the
photons to be completely random process, we have for the average
square of the momentum change of the ensemble of atoms~\cite{Minogin}:
\begin{equation}
\langle\Delta {p}_{x}^{2}\rangle =\langle\Delta {p}_{0x}^{2}\rangle+
\hbar^{2}{k}^{2}\langle N_{s}\rangle+
\hbar^{2}{k}^{2}\langle\cos^{2}\theta\rangle\langle N_{s}\rangle.
\label{eq:DpDpx} 
\end{equation}
The first term on the right hand side of \eqref{eq:DpDpx} is the
initial moment distribution of the atoms, the second is due to the
stimulated processes (absorption and emission), the third term is due
to the fluctuations of the momentum in the spontaneous emission of
photons. This equation is the basis for the computer simulation of the
momentum diffusion in the field of one travelling wave.  According to
\eqref{eq:DpDpx}, for each random change of the atomic momentum due to
the spontaneous emission of light, one change in the atomic momentum
at $ \pm\hbar k $ occurs due to the stimulated processes. This
algorithm for calculating momentum diffusion of atoms is valid for a
weak field $ \Omega_{0} \lesssim \gamma $.  If the intensity of the
counter-propagating waves is large, we will take into account the
momentum diffusion due to stimulated processes in the same way,
knowing that the results we have obtained is the approximation to the
strict consideration.  Since we suppose the motion of atoms for the
same intensity of criss-cross waves, one should expect that the
fluctuation change of the square of $ x $ and the $ y $-components of
the momentum of the atoms due to stimulated processes are the same for
averaging after several modulation periods.

After each step $ \Delta t $ of the integration of the equations (\ref
{eq:ma}), (\ref{eq:dci}) and (\ref{eq:dcii}) we check if a quantum
jump occured during it and normalize the state vector. If the quantum
jump occured, the atom's velocity along the $ Ox $ axis varies by
\begin{eqnarray}
  \Delta{}v_{x}&=&\hbar{}k\mathop{\mathrm{sgn}}(\epsilon_1-0.5)\frac{1+\mathop{\mathrm{sgn}}(\epsilon_2-0.5)}{2m}
                   +\hbar{}k\mathop{\mathrm{sgn}}(\epsilon_3-0.5)\frac{1+\mathop{\mathrm{sgn}}(\epsilon_4-0.5)}{2m},
  \label{eq:dvx}\\
  \Delta{}v_{y}&=&\hbar{}k\mathop{\mathrm{sgn}}(\epsilon_1-0.5)\frac{1-\mathop{\mathrm{sgn}}(\epsilon_2-0.5)}{2m}
                   +\hbar{}k\mathop{\mathrm{sgn}}(\epsilon_3-0.5)\frac{1-\mathop{\mathrm{sgn}}(\epsilon_4-0.5)}{2m},
  \label{eq:dvy}
\end{eqnarray}
where $ \epsilon_{ 1,2,3,4} $ are the random numbers that are
uniformly distributed in the interval $[0,1]$.  Here two terms
simulate the fluctuation of the momentum with the spontaneous emission
of the photon along the positive and negative directions of each of
the axes; another two simulate the fluctuation of the momentum in the
same directions due to the fluctuation of the stimulated absorption
and emission processes.

\section{Results of the numerical simulatuon}
\label{sec:--Results}

The time evolution of an atomic ensemble in a bichromatic field is
determined by the parameters of the interaction of atoms with the
field and the initial conditions. To simplify the analysis, we do not
take into account the influence of the initial atomic distribution
function of the coordinates and velocity components. We assume that
the atoms begin to move from the same point, which is the origin of
the coordinates ($ x_ 0 = 0 $, $ y_{0} = 0 $), with the same velocity.
We carried out our calculations for the atom $^{23}\mathrm {Na}$, for
which a cyclic interaction with the field can be
organized~\cite{Metcalf}. The wavelength of the transition
$ 3{}^{2}S_{1/2} - 3 {}^{2} P_{3/2}$ is $ \lambda = 589.16 $ nm, the
spontaneous emission rate $ \gamma = 2 \pi \times10 $~MHz, the Doppler
cooling limit for sodium atoms is $T_{D} = 240$~$\mu{}$K
\cite{Metcalf}.  Similar calculations for the case of one standing
bichromatic wave were carried out earlier in the
paper~\cite{Romanenko2016}.  We consider the motion of atoms in the
case of criss-cross bichromatic waves for parameters similar to those
used for numerical simulation in the paper~\cite{Romanenko2016},
focusing on those that correspond to the formation of traps for atoms
and the formation of periodic structures. In our case, obviously, one
should expect the formation of a two-dimensional trap and
two-dimensional spatial structures.

\begin{figure}[h]
  \centerline{\includegraphics{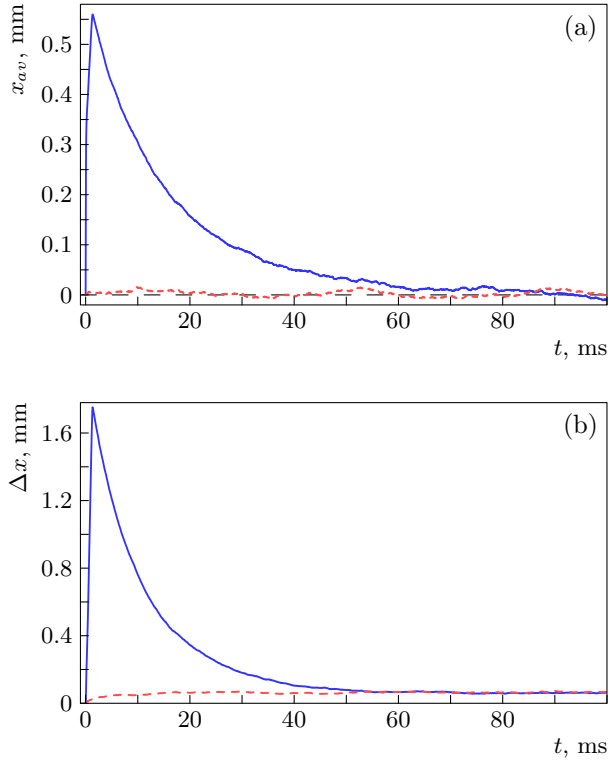}}
  \caption{The time dependence of the mean coordinate $ x_{av} $ (a)
    and the root mean square deviation $\Delta x $ (b) of 100
    $^{23}\mathrm{Na} $ atoms under their interaction with criss-cross
    standing bichromatic waves. The components of the initial velocity
    of atoms are $ v_{0x} = v_{y0} = 5 $~m/s (solid curve),
    $ v_ {0x} = v_ {y0} = 0 $ (dashed curve), $ \Omega = 200 $~MHz,
    $ \delta = 2\pi\times20 $~MHz
    ($ \delta_ {1} = \delta_ {2} = \delta_ {5} = \delta_ {6} =
    2\pi\times110$~MHz,
    $ \delta_{3} = \delta_{4} \fh{=} \delta_{7} = \delta_{8} \fh{=}
    -2\pi\times90 $~MHz),
    the Rabi frequencies are the same and equal to
    $ \Omega_{0} = 2 \pi \times100 $~MHz, the initial phases of the
    waves equal to zero. Before the interaction with the field the
    atoms occupy the ground state.}
\label{fig-2}
\end{figure}

To check the possibility of forming a two-dimensional trap by the
field of criss-cross bichromatic waves, the time dependences of the
center of mass (mean values of abscissa and ordinates, $x_{av}$ and
$y_{av}$) of the ensemble of atoms and the root-mean-square deviation from the
mean values $ \Delta x $, $ \Delta y $ were calculated. An example
of these dependencies is shown in Fig.~\ref{fig-2}.
We do not present here the corresponding time dependencies of
ordinates; they are similar to the corresponding dependences of
$ x_{av} $ and $ \Delta x $.  As we see, in the case of the initial
velocity of the atoms $ v_{0x} = v_{0} = 5 $~m/s, when atoms move from
the origin of coordinates, they are subjected to the light pressure
force in the direction to the point $ x = 0, y = 0 $.  Subsequently,
the value $ x_ {av} $ fluctuates near $ x = 0 $ (the same applies to
$ y_ {av} $), and $ \Delta x $ {and $ \Delta y $} fluctuates near
$ \Delta x \approx66 $~$\mu$m.  For greater certainty that
approximately the stationary value is achieved, we carried out the
similar calculations for zero initial velocity. The result was the
same (see dashed curve in Fig.~\ref{fig-2}).  This is a little more
than a similar result for the one-dimensional trap
$ \Delta x \approx50 $~$\mu$m (see Fig. 8 in
paper~\cite{Romanenko2016}). We should note that in order to have a
coordinate-dependent light pressure force directed towards the
coordinates origin, the intensity of the light waves should be quite
large~\cite{Romanenko2016}.

\begin{figure}[h]
 \centerline{\includegraphics{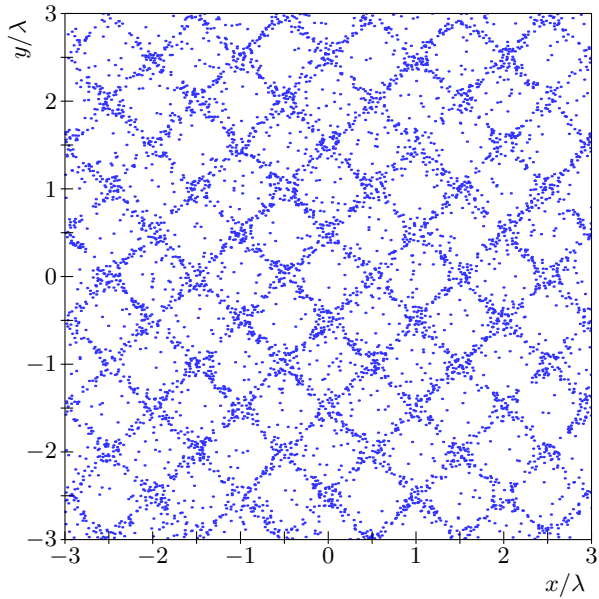}}
 \caption{A fragment of the spatial distribution of 50,000
   $^{23}\mathrm{Na} $ atoms after $100\,\mu$s of the interaction with
   the criss-cross standing bichromatic waves. The initial velocity of
   the atoms is zero, $ \Omega = 2\pi\times40 $~MHz,
   $ \delta = 2\pi\times20$~MHz
   ($ \delta_ {1} = \delta_{2} = \delta_{5} = \delta_{6} =
   2\pi\times120$~MHz,
   $ \delta_{3} = \delta_{4} = \delta_{7} = \delta_{8} = -
   2\pi\times80$~MHz), the Rabi frequencies of the waves are the same
   and equal to $ \Omega_ {0} = 2 \pi \times100 $~MHz, the initial
   phases of all waves are equal to zero.}
\label{fig-3}
\end{figure}

Now we consider the formation of spatial structures of atoms by the
field of the criss-cross standing bichromatic waves.  We remind that
the field of a standing bichromatic wave can form a one-dimensional
grating of atoms with their grouping in planes at distances
$ \frac {1} {4} \lambda + \frac {1} {2} n \lambda $ from the origin of
coordinates, where $ n $ is an arbitrary integer~\cite{Romanenko2016}.
Fig.~\ref{fig-3} depicts a fragment of the spatial distribution of
$^{23}\mathrm{Na}$ atoms after $ 100\,\mu$s of their interaction with
the criss-cross stationary bichromatic waves.
During this time, the stationary value of the components of the root
mean square deviation of the atoms velocity from the average value
$\Delta v_{x}=\Delta v_{y}\fh{=}0.248$~m/s is established. For
$t=100$~$\mu$s, the root mean square deviation of the atoms
coordinates from the mean value is
$ \Delta{x} \approx \Delta{y} \approx4.5$~$\mu$m and continues to
grow.

\begin{figure}[h!]
  \centerline{\includegraphics{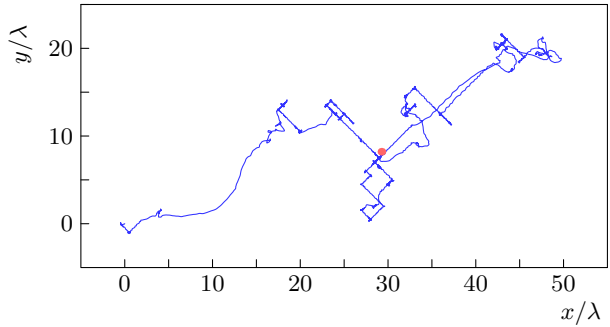}}
  \caption{Trajectory of the motion of one of the atoms for the
    parameters corresponding to Fig.~\ref{fig-3}. Time of motion is
    500~$\mu$s. The red circle shows the point corresponding to time
    500~$\mu$s}
\label{fig-4}
\end{figure}

\begin{figure}[h!]
\centerline{\includegraphics{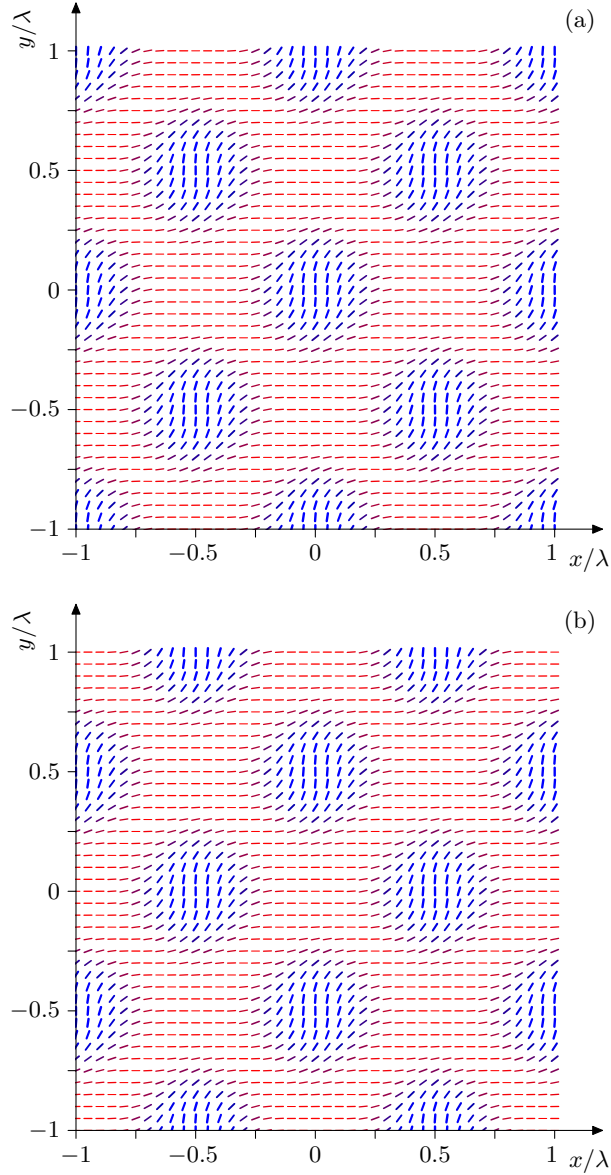}}
\caption{The distribution of the energy density of the electric field
  of the criss-cross standing bichromatic waves for cases $ \varphi = 0 $
  (a), when the phases of all waves are the same, and  
  $ \varphi = \pi $ (b), when the phases of the waves propagating
  along the ordinate axis differ on $ \varphi = \pi $ from the phases
  of the waves propagating along the abscissa axis.  The angle of
  inclination of the strokes to the abscissa axis is proportional to
  the energy density of the field.}
\label{fig-5}
\end{figure}

\begin{figure}[h!]
    \centerline{\includegraphics{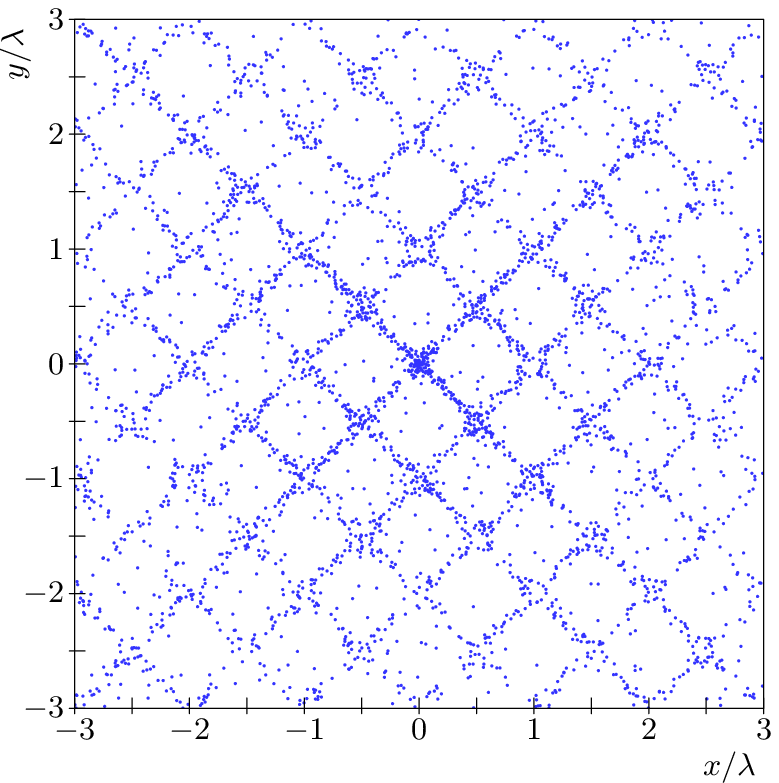}}
    \caption{A fragment of the spatial distribution of 50,000 atoms
      $^{23}\mathrm{Na} $ after $100\mu$ of interaction with the
      criss-cross standing bichromatic waves. The initial velocity of
      atoms is zero, $\Omega \fh{=} 2\pi\times40$~MHz,
      $\delta = 2\pi\times20 $~MHz
      ($\delta_{1}=\delta_{2}=\delta_{5}=\delta_{6}=2\pi\times120$~MHz,
      $\delta_{3}=\delta_{4}\fh{=}\delta_{7}=\delta_{8}=-2\pi\times80$~MHz),
      the Rabi frequencies of the waves are the same and equal to
      $ \Omega_{0} = 2\pi \times100 $~MHz,
      $ \varphi_{1} = \varphi_{2} = \varphi_{3} = \varphi_{4} = 0 $,
      $ \varphi_{5} = \varphi_{6} = \varphi_{7} = \varphi_{8} = \pi$. }
\label{fig-6}
\end{figure}

The spatial distribution of atoms depicted in Fig.~\ref{fig-3} is
related to the spatial distribution of the the density of the energy
of the electric field, averaged over time, which in this case is
described by the expression
\begin{equation}
  \label{eq:fieldpower0}
  w=8\varepsilon_{0}E_{0}^{2}
  \left[
    \sin^{2}(\pi x/\lambda)+\sin^{2}(\pi y/\lambda)-1
  \right]^{2}.
\end{equation}
Here we have taken into account that we are considering the distances
of atoms from the origin of the coordinates, for which
$ x \Delta k \ll1 $, $ y \Delta k \ll1 $. From the
expression~\eqref{eq:fieldpower0} it is easy to see that the energy
density is zero along the families of straight lines described by the
equations
\begin{eqnarray}
y&=&x+\frac{\lambda}{2}+n_{1}\lambda,
  \label{eq:line1}\\
y&=&-x+\frac{\lambda}{2}+n_{2}\lambda,
    \label{eq:line2}
\end{eqnarray}
where $ n_{1} $, $ n_{2}$ are integers.  As shown in Fig.~\ref{fig-3},
most of the atoms are near the straight lines described by the
equations~\eqref{eq:line1}, \eqref{eq:line2}, that is, at points with
low density of energy of the electric field.  The explanation for such
a phenomenon is quite simple. If the atom, while moving in the weak field,
moves almost along the straight lines \eqref{eq:line1},
\eqref{eq:line2}, the direction of it's velocity changes with a very
small probability, or changes after reaching the points with strong
field. If the atoms pass through the region of a strong field, the
probability of changing the direction of their velocity is much
higher. Thus, in the distribution of velocities for atoms, we have a
maximum near the directions corresponding to the motion of atoms along
the straight 1lines \eqref{eq:line1}, \eqref{eq:line2}, which explains
the formation of the spatial grating shown in Fig.~\ref{fig-3}.
Figure~\ref{fig-4} shows the trajectory of one of the atoms. 
It is well visible that a large part of the trajectory consists of
segments directed at the angle $\pm\frac{1}{4}\pi$ to the axis of
abscissa and ordinate in accordance with the equations
\eqref{eq:line1}, \eqref{eq:line2}.  Consequently, the two-dimensional
grating of atoms in Fig.~\ref{fig-3} has a dynamic character.  The
atoms move most of the time close to directions of the straight lines,
defined by the equations \eqref{eq:line1} and \eqref{eq:line2}, from
time moving from one straight line to another.  Figure \ref{fig-5}
illustrates the distribution of the energy density of the electric
field for the cases $ \varphi = 0 $ (a) and $ \varphi = \pi $ (b).
It is seen that energy is mainly concentrated near the points
$ x = n_{x} \lambda $, $ y = n_{y} \lambda $ (a), where $ n_{x} $,
$ n_{y} $ are arbitrary integers, and near the points
$ x = n_{x} \lambda + \frac{1}{2}\lambda $,
$ y = n_{y} \lambda + \frac {1} {2} \lambda $~(b).  The structure
formed by atoms for $ \varphi {} = 0 $ (Fig.~\ref{fig-3}) corresponds
to the regions of the weak field in Fig.~\ref{fig-5}a.  If
$ \varphi = \pi $, the areas of the strong field are shifted by
$ \lambda / 2 $ along the axis of abscissas and ordinates in
comparison with the case $ \varphi = 0$; we see again that most of the
atoms are in the region of a weak field (see Fig.~\ref{fig-6}).
Contrary to Fig.~\ref{fig-3}, in Fig.~\ref{fig-6} there is a
noticeable difference in the density of atoms in the center of the
figure and at its periphery.  This is due to the slower expansion of
the cloud of atoms, since the field is very small near the initial
coordinates ($x_{0}=0$, $y_{0}=0$) for $\varphi=\pi$. As the result
there is a smaller characteristic dimension of the atomic cloud
($\Delta x = \Delta y = 2.5 $~$ \mu$m versus
$\Delta x=\Delta y=4.5$~$\mu$m for $\varphi=0$).

Note that in the general case of the arbitrary values of the phase
shift $\varphi$ between the criss-cross waves, generally speaking,
there are no such lines in space, for which the density of the
electric field energy, averaged over time,
\begin{eqnarray}
  \label{eq:field_power}
  w&=&4\varepsilon_{0}E_{0}^{2}
       \left[ 2
       \left(
        \sin^{2}(\pi x/\lambda)+\sin^{2}(\pi y/\lambda)
       \right)^{2}+\right.
       \nonumber\\
       &&\mbox{}+4(\cos\varphi-1)\sin^{2}(\pi x/\lambda)\sin^{2}(\pi y/\lambda)+\nonumber\\
&&
   \mbox{}+\left(1-
        2\sin^{2}(\pi x/\lambda)-2\sin^{2}(\pi y/\lambda)\right)\times\nonumber\\
&&\left.\mbox{}\times(1+\cos\varphi)\right],
\end{eqnarray}
would be zero. For example, for $\varphi=\pi/2$ the
equation~\eqref{eq:field_power} leads to the expression
\begin{eqnarray}
  w&=&4\varepsilon_{0}E_{0}^{2}
  \left[
    \sin^{2}(\pi x/\lambda)+\sin^{2}(\pi y/\lambda)-1
  \right]^{2}+\nonumber\\
&&\mbox{}+\sin^4(\pi x/\lambda)+\sin^{4}(\pi y/\lambda),
\end{eqnarray}
which at any point is not zero.  At the same time there may be
rectilinear regions with a rather low energy density, which makes it
possible to form spatial gratings. In particular, the results, which
are close to the results shown in Fig. \ref{fig-3} and \ref{fig-6},
are observed at small changes in the phase shift between the
criss-cross waves, respectively, at $\varphi=0.05\pi$ and $\varphi=0.95\pi$.

The spatial structures of the clouds of atoms in the fields of the
criss-cross standing bichromatic waves arise due to the joint action
of both waves on the atom and can not be interpreted by generalizing
the results of the study of spatial structures of atoms in the field
of a standing bichromatic wave~\cite{Romanenko2016} to a
two-dimensional case.  Indeed, in one standing bichromatic wave, atoms
are grouped in planes with
$ x = \frac{1}{4} \lambda + \frac {1}{2} n \lambda $ ($ n $ is an
integer) for waves propagating along the abscissa axis and
$ y = \frac{1}{4} \lambda + \frac{1}{2} n \lambda $ for waves
propagating along the ordinate axis.

Extrapolating these results in a two-dimensional case, one would
expect the grouping of atoms near the points with coordinates
$ x = \frac{1}{4}\lambda + \frac{1}{2}n_{x}\lambda $,
$ y = \frac{1}{4}\lambda + \frac{1}{2}n_{y}\lambda $, which does not
correspond to the results obtained in this paper, namely, a spatial
grating with the period $\frac{\sqrt{2}}{2}\lambda $, oriented at an
angle $\pi/4 $ to the abscissa and ordinate axes.

\section{Conclusions}

We investigated the spatial distribution of atoms in the field of
criss-cross bichromatic standing waves.  As it turned out, using the
previous study of the spatial distribution of atoms in the field of a
standing bichromatic wave~\cite{Romanenko2016}, we can estimate the
parameters and character of the distribution of atoms in the field of
criss-cross standing standing bichromatic waves only partially.  Thus,
the parameters of the distribution of atoms in the light trap created
by such fields are in good agreement with the parameters expected on
the basis of paper~\cite{Romanenko2016}.  At the same time, the
appearance of the spatial structure can not even be approximated by
extrapolating the results of the paper~\cite{Romanenko2016} for a
one-dimensional case on a two-dimensional case; instead of the
expected grouping of atoms near the points
$ x = \frac{1}{4}\lambda\fh{+}\frac{1}{2} n_{x} \lambda $,
$ y = \frac{1}{4}\lambda+\frac{1}{2} n_{y} \lambda $, where $ n_{x}$,
$ n_{y} $--- arbitrary integers, the atoms group in a neighborhood of
lines that correspond to the minimum energy. As a result, the cloud of
atoms form a grid with the side equal to $\frac{\sqrt{2}}{2}\lambda$.
This grating has a dynamic character. The atoms, that are part of it,
spend some, relatively short, time there, moving along its sides, then
exit and after some time enter another place of the grating.

\begin{acknowledgments}
  The publication is based on the research supported by the
  goal-oriented complex program of fundamental researches of the
  National Academy of Sciences of Ukraine ``Fundamental issues in
  creation of new nanomaterials and nanotechnologies'' (grant
  No. 3/18-H).
  \end{acknowledgments}
\bibliography{Bichromatic-2D-ArXiv}
\end{document}